# Relation between rare-earth magnetism and the magnetocaloric effect in multiferroic hexagonal manganites

R. Dragland[1], C. Salazar Mejía[2], I. Hansen[1], Y. Hamasaki[3], E. C. Panduro[1],
Y. Ehara[4], T. Gottschall[2], D. Meier[1,*], J. Schultheiß[1,5,*]

[1] Department of Materials Science and Engineering, Norwegian University of Science and Technology (NTNU), 7034, Trondheim, Norway
[2] Dresden High Magnetic Field Laboratory (HLD-EMFL), Helmholtz-Zentrum Dresden-Rossendorf, 01328 Dresden, Germany
[3] Department of Applied Physics, National Defense Academy, Hashirimizu, Yokosuka 239-8686, Japan
[4] Department of Communications Engineering, National Defense Academy, Hashirimizu, Yokosuka 239-8686, Japan
[5] Department of Mechanical Engineering, University of Canterbury, 8140 Christchurch, New Zealand
* dennis.meier@ntnu.no, jan.schultheiss@ntnu.no

The magnetocaloric effect enables magnetic refrigeration and plays an important role for cooling at cryogenic temperatures, which is essential for emergent technologies such as hydrogen liquefaction and quantum computing. Here, we study the magnetocaloric effect in multiferroic hexagonal manganites by conducting direct adiabatic temperature measurements in pulsed magnetic fields exceeding 20 T. Data gained on polycrystalline HoMnO$_3$, ErMnO$_3$, TmMnO$_3$, and YMnO$_3$ demonstrate a direct correlation between the magnetic 4f-moments and the measured adiabatic temperature change. In HoMnO$_3$, i.e., the system with the largest magnetic 4f-moments, significant temperature changes, $\Delta T_{ad}$, of up to 20.1 K are observed, whereas the effect is largely suppressed in YMnO$_3$. Our systematic investigations show the importance of the rare-earth magnetism for the magnetocaloric effect in multiferroic hexagonal manganites at cryogenic temperatures, reaching about 64% of the adiabatic temperature changes reported for gadolinium at room temperature.

## I. Introduction

Cryogenic refrigeration is essential for numerous advanced technologies, ranging from magnetic resonance imaging and particle accelerators for high-energy physics to quantum computing. Furthermore, the ongoing energy transition, with hydrogen gas as a green energy vector, has renewed the interest in cryogenic refrigeration, [1,2] particularly for developing next-generation hydrogen gas liquefiers that can efficiently reach the required temperatures below 21.1 K. A major challenge is that conventional compression-based cooling methods become inefficient and expensive at these low temperatures, [3,4] which makes a development of novel refrigeration approaches highly desirable.

A promising alternative for efficient and sustainable cooling at cryogenic temperatures is the caloric effect, utilizing field-induced order-disorder phase transitions under adiabatic conditions. Adiabatic temperature changes originating from changes in electric [5,6] or elastic [7,8] order, however, decrease towards cryogenic temperatures and face fatigue challenges, respectively. Thus, recent research has focused on magnetic materials, [9,10] particularly binary and ternary intermetallic compounds. Examples include rare-earth nickelates, [11] aluminides, [12] and Laves phases, [13] which show suitable magnetic-field-induced phase transitions in the cryogenic temperature regime and large adiabatic temperature changes. Also non-metals, such as rare-earth borides, show a promising cooling performance and their physical properties have been studied in detail. [14] In comparison, much less is known about the magnetocaloric properties in rare-earth oxides and much of the underlying physics remain to be explored.

Rare-earth oxides are interesting for magnetocaloric cooling applications because of their complex magnetism. For example, driven by the interaction of coexisting magnetic 4f- and 3d-moments, unusual frustrated states, [15,16] long-range modulated spin textures, [17] and an anisotropic magnetic behavior arise, [18] providing new degrees of freedom for material design. [9] The rare-earth hexagonal manganites, $R$MnO$_3$ ($R$ = Sc, Y, In, or Dy-Lu), have been investigated intensively with respect to their magnetism and multiferroic properties, [19-27] and their magnetic phase diagrams at cryogenic temperature are well-understood [28-31]. Because of this, the hexagonal $R$MnO$_3$ family is an ideal model system for the study of magnetic refrigeration in rare-earth oxides and application opportunities for cryogenic cooling. In the $R$MnO$_3$ systems, antiferromagnetic ordering of the Mn$^{3+}$ 3d-moments occurs at $T_N^{Mn^{3+}} \lesssim 120$ K, whereas the $R^{3+}$ 4f-moments exhibit spontaneous long-range order at $T_N^{R^{3+}} \lesssim 10$ K. Regarding refrigeration, the magnetocaloric effect in hexagonal rare-earth manganites has already been studied at room-temperature. [32,33] In addition, various studies have reported promising magnetocaloric properties in the cryogenic temperature regime. [34-42] The research primarily focused on magnetically anisotropic single crystals, which allow for caloric cooling under a constant magnetic field and demonstrated the potential of $R$MnO$_3$ systems for caloric cooling. The microscopic origin of the magnetocaloric effect at cryogenic temperatures, however, and the role the magnetic 3d- and 4f-moments play for the observed cooling, remain to be clarified.

Here, we measure adiabatic temperature changes at cryogenic temperature for selected polycrystalline hexagonal manganites with varying magnetic 4f-moments (i.e., HoMnO$_3$, ErMnO$_3$, TmMnO$_3$, and YMnO$_3$). Specific heat measurements show that our polycrystals exhibit the established material-specific sequences of magnetic phase transitions. Direct adiabatic temperature change measurements under pulsed magnetic fields up to 22.3 T verify a pronounced magnetocaloric response. We find substantial adiabatic temperature changes up to $\Delta T_{ad}$ = 20.1 K in HoMnO$_3$ and less pronounced effects in ErMnO$_3$ and TmMnO$_3$, which we attribute to their smaller magnetic 4f-moments. The latter is corroborated by the suppression of adiabatic temperature changes in YMnO$_3$, demonstrating that the magnetocaloric effect in hexagonal rare-earth manganites at cryogenic temperatures is predominantly driven by the magnetic order of the rare-earth sublattice.

## II. Experimental Procedures
### A. Synthesis

The polycrystals of HoMnO$_3$, ErMnO$_3$, TmMnO$_3$, and YMnO$_3$ used in this work were synthesized via a solid-state synthesis approach from raw-oxide materials. For the synthesis, Tm$_2$O$_3$ (99.9% purity, AlfaAesar, Haverhill, MA, USA), Ho$_2$O$_3$ (99.9% purity, Merck, Darmstadt, Germany), Er$_2$O$_3$ (99.9% purity, Alfa Aesar), Y$_2$O$_3$ (99.9% purity, Alfa Aesar), and Mn$_2$O$_3$ (99.0%, Alfa Aesar) were utilized. Tm$_2$O$_3$, Ho$_2$O$_3$, Er$_2$O$_3$, and Y$_2$O$_3$ were furnace dried at 900°C, while Mn$_2$O$_3$ was dried at 700°C for a dwell time of 12 hrs. Stoichiometric amounts of both raw powders were weighed, ball milled for 24 h at 205 rpm, and heat treated at 1100°C for 12 hrs in air. The powder was pressed uniaxially into cylindrical pellets with a diameter of 10 mm, followed by isostatic compaction under a pressure of 200 MPa. The final densification of the pressed pellet was done at 1400°C for 12 hrs. A more detailed description of the synthesis can be found in ref. [43] for the case of ErMnO$_3$.

### B. Crystallographic and microstructural characterization

The processed ceramic pellets were investigated by XRD (XRD, D8 ADVANCE, Bruker, Billerica, MA, USA), and the respective lattice parameters were determined by Pawley fitting [44] using Topas. For microstructural characterization, samples were lapped utilizing a 9-µm grained Al$_2$O$_3$ water suspension (Logitech, Glasgow, Scotland), followed by a subsequent polishing step utilizing a silica slurry (SF1 Polishing Fluid, Logitech). Scanning electron microscopy (SEM) was performed on a Helios NanoLab DualBeam Focused Ion Beam (Thermo Fischer Scientific, MA, USA) using secondary electrons for imaging. The electron beam had a 5 kV acceleration voltage and a 0.17 nA beam current. Piezoresponse force microscopy (PFM) was performed using an NT-MDT system (NT-MDT, Moscow, Russia) with an electrically conductive tip (Asyelec.01-R2, Oxford Instruments, Abingdon, UK). The PFM measurement was performed at a frequency of 40.14 Hz with a peak-to-peak voltage of 10 V. The vertical deflection of the laser was read out as $R \cdot \cos\Phi$, with amplitude, $R$, and the phase $\phi$ of the piezoresponse using a lock-in amplifier (SR830, Stanford Research Systems, CA, USA).

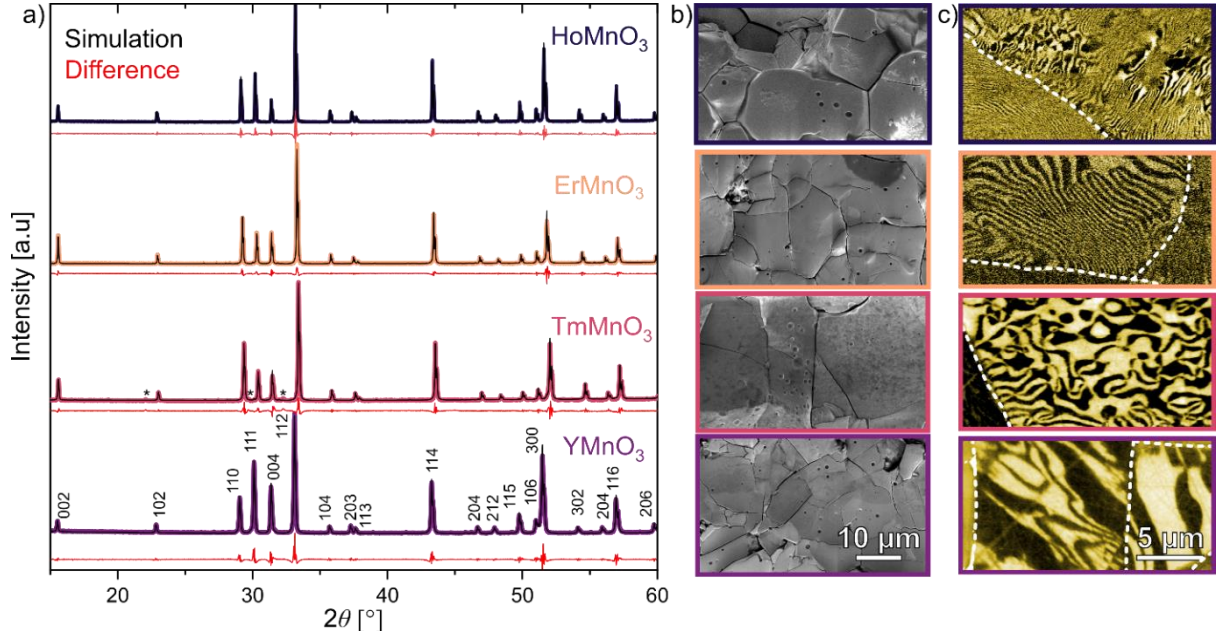

**Fig. 1:** Crystal- and microstructure characterization of polycrystalline hexagonal manganites. a) XRD diffraction data of polycrystalline pellets. A Pawley fit (black curve) is fitted to the experimental data, with the difference displayed in red. The data verifies the hexagonal target phase, space group $P6_3cm$. A minor presence of secondary phases can be identified in the diffraction pattern of TmMnO$_3$, as indicated by an asterisk (*). b) SEM micrograph of the polished surface of the polycrystalline pellets. c) Vertical component of the PFM response of an area featuring several grains. The boundary between the grains is displayed by a dashed white line.

### C. Caloric measurements

The heat capacity of the polycrystals was measured as a function of temperature in constant magnetic fields of 0 T, 2 T, 5 T, and 9 T using a Dynacool PPMS system (Quantum Design, San Diego, CA, USA). Adiabatic temperatures changes were measured directly at the Dresden High Magnetic Field Laboratory of the Helmholtz-Zentrum Dresden-Rossendorf in pulsed magnetic fields up to 22.3 T between 10 K and 100 K. The time to reach the maximum magnetic field was ~17 ms. To realize direct temperature measurements, each sample was cut into two flat rectangular pieces, approximately 4 x 2 x 1 mm$^3$, utilizing a diamond wire saw (Well Diamond Wire Saws SA, Mannheim, Germany) and a differential type-E thermocouple made from 25 µm thin constantan and chromel wires was fixed between these pieces by a small amount of silver epoxy in a sandwich-like structure to measure the adiabatic temperature changes directly. After the samples were fixed on the sample holder, the insert was evacuated to a vacuum of 10$^{-3}$ mbar. More details on the direct adiabatic temperature change measurements can be found in ref. [45].

### III. Results
#### A. Crystallographic and microstructural characterization

We begin our analysis by characterizing the crystal structure of our polycrystalline samples. Corresponding XRD data are displayed in Fig. 1a, showing that all samples exhibit $P6_3cm$ space group symmetry as expected for hexagonal manganites. Only in the case of TmMnO$_3$, minor secondary phases are observed as indicated by an asterisk (*), which we attribute to unreacted starting materials. The lattice parameters extracted from a Pawley fit of the XRD data, along with the calculated unit cell volume, are summarized in Tab. 1. A clear correlation is observed between the unit cell volume and the size of the rare-earth atom consistent with literature [19].

Microstructural analysis conducted via SEM, Fig. 1b, confirms the high density and polycrystalline nature of the samples. PFM

**Tab. 1.** Crystallographic characterization and magnetic transition temperatures. Unit cell parameters *a* and *c* of the $P6_3cm$ space group, along with the resulting unit cell volume, *V*, are extracted from the XRD data shown in Fig. 1a) using Pawley fits. TmMnO$_3$, the material with the smallest ionic radius, has the smallest unit cell volume. The unit cell volume increases with increasing ionic radius. The ordering temperatures related to the Mn$^{3+}$ and $R^{3+}$ ions, extracted from specific heat data (Fig. 2a-d), are displayed.

| Composition | Ionic radius of $R^{3+}$ (Å) | $a$ (Å) | $c$ (Å) | $V$ (Å$^3$) | $T_N^{Mn^{3+}}$ (K) | $T_N^{R^{3+}}$ (K) |
|---|---|---|---|---|---|---|
| HoMnO$_3$ | 0.901 | 6.140(0) | 11.415(4) | 372.7 | 74 | 5 |
| ErMnO$_3$ | 0.890 | 6.115(2) | 11.408(1) | 369.5 | 79 | 4 |
| TmMnO$_3$ | 0.880 | 6.089(0) | 11.383(8) | 365.5 | 81 | 10 |
| YMnO$_3$ | 0.900 | 6.146(0) | 11.401(8) | 373.0 | 73 | - |

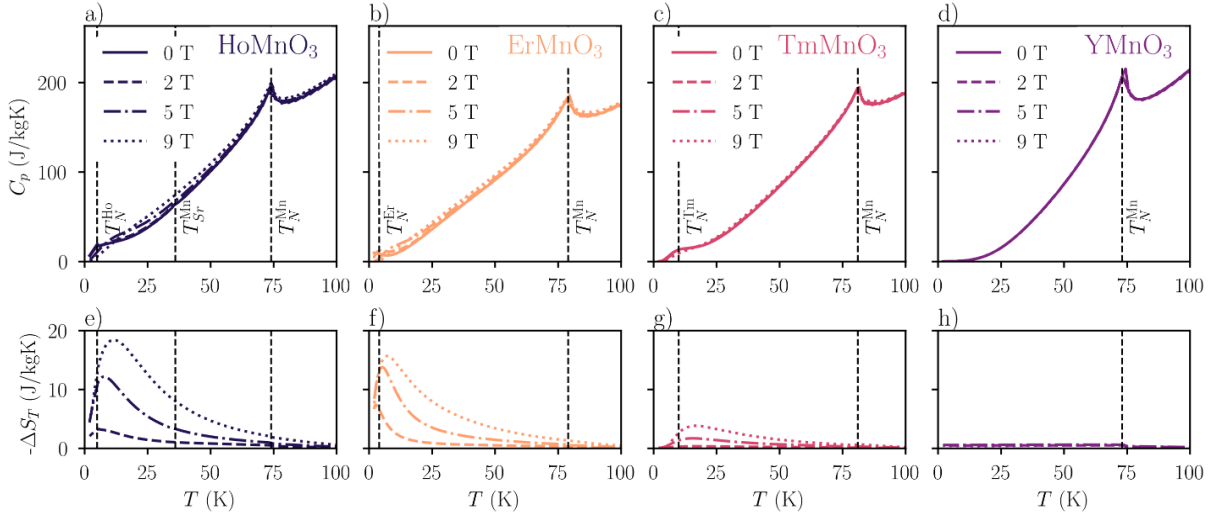

**Fig. 2.** Specific heat and entropy data of polycrystalline hexagonal manganites. Temperature-dependent specific heat (cooling cycle only) measurements on a) HoMnO$_3$, b) ErMnO$_3$, c) TmMnO$_3$, and d) YMnO$_3$ under magnetic fields of 0 T, 2 T, 5 T, and 9 T. Dashed vertical lines display the ordering temperature of the manganese sublattice, $T_N^{Mn^{3+}}$, the reorientation of the magnetic moment of the manganese, $T_{SR}^{Mn^{3+}}$, and the ordering of the rare-earth sublattice, $T_N^{R^{3+}}$. The entropy change derived from the specific heat data under different magnetic fields, according to eq. 1, is displayed in panels e-h).

scans, displayed in Fig. 1c, show the established ferroelectric domain structure composed of vortex- and strain-driven stripe-like domains as discussed elsewhere [46-48]. The data extends previous investigations on polycrystalline hexagonal ErMnO$_3$ [43,49-51] and DyMnO$_3$ [52] to isostructural HoMnO$_3$, TmMnO$_3$, and YMnO$_3$ and confirms that these compounds exhibit the $R$MnO$_3$-characteristic structural and ferroelectric properties.

**B. Specific heat**

After verifying the crystallographic symmetry and the microstructure of our polycrystalline samples, specific heat measurements are conducted to confirm that the materials display the expected magnetic phase transitions at cryogenic temperatures. The specific heat data over a temperature range from 2 K to 100 K are presented in Fig. 2a-d, showing different anomalies that can be attributed to magnetic phase transitions associated with the Mn$^{3+}$ and $R^{3+}$ sub-lattices. Consistent with previous studies on the hexagonal manganites, [53-55] a peak is observed in the specific heat data obtained during cooling below 100 K as indicated by the dashed lines. This peak occurs at the Néel temperature, $T_N^{Mn^{3+}}$, indicating the onset of antiferromagnetic ordering of the magnetic 3d-moments of the Mn$^{3+}$ ions, forming frustrated triangular spin arrangements in the $ab$-plane as discussed, e.g., in refs. [27,56]. The respective transition temperatures are summarized in Tab. 1, showing a good agreement with the previously reported transition temperatures: ~76 K for HoMnO$_3$, [19] ~79 K for ErMnO$_3$, [19] ~84 K for TmMnO$_3$ [28] and 72 K for YMnO$_3$ [55]. As the temperature is further reduced, we observe another transition in HoMnO$_3$ at $T_{SR}^{Ho^{3+}}$ = 36 K, which originates from a spin reorientation (SR) of the magnetic 3d-moments. [53] Anomalies in the specific heat data below 10 K are associated with spontaneous long-range ordering of the magnetic 4f-moments on the $R^{3+}$ sites; we find the lowest transition temperature for Er$^{3+}$ ($T_N^{Er^{3+}}$ = 4 K) and the highest for Tm$^{3+}$ ($T_N^{Tm^{3+}}$ = 10 K) (Tab. 1), which is in reasonable agreement with literature data (~3 to 5 K [26,27]). Note that Y$^{3+}$ has no magnetic 4f-moment and, hence, does not exhibit an additional magnetic phase transition below 10 K. In summary, the specific heat data confirms that the polycrystals synthesized for this work exhibit the established sequence of magnetic phase transitions, which originate from the onset of spontaneous long-range antiferromagnetic ordering of the magnetic moments on the Mn$^{3+}$ and $R^{3+}$ sites.

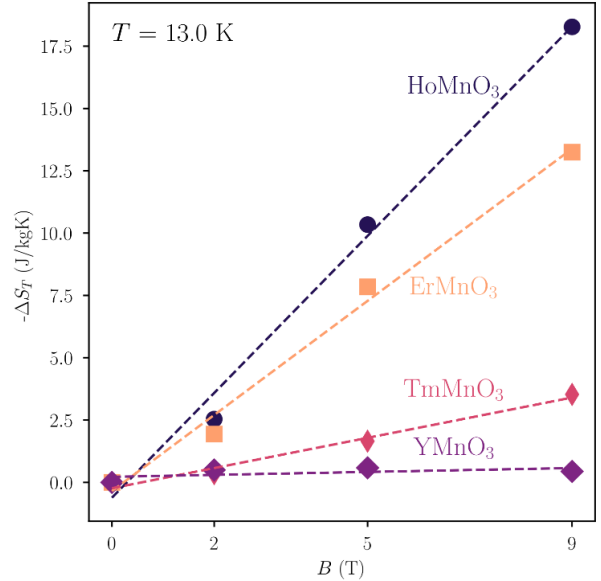

**Fig. 3.** Magnetic-field dependence of the entropy change at $T$ = 13 K. The dashed lines are a guide to the eye.

**Tab. 2.** Entropy and adiabatic temperature changes. The calculated effective magnetic moment, $\mu_B$ for different $R^{3+}$ ions is displayed. The entropy change is calculated from specific heat data according to Eq. 1 as displayed in Fig 2e)-h). Adiabatic temperature changes are measured directly and displayed in Fig. 4.

| Composition | Effective magnetic moment ($\mu_B$) of $R^{3+}$ | $B$ = 5 T | | $B$ = 10.8 T | $B$ = 22.3 T |
|---|---|---|---|---|---|
| | | $\Delta S_T$ [JKg$^{-1}$K$^{-1}$] | $\Delta T_{ad}$ [K] | $\Delta T_{ad}$ [K] | $\Delta T_{ad}$ [K] |
| TmMnO$_3$ | 7.6 | 2.1 | 1.9 | 5.0 | 8.6 |
| ErMnO$_3$ | 9.6 | 5.5 | 5.9 | 11.1 | 15.1 |
| HoMnO$_3$ | 10.6 | 10.5 | 5.2 | 12.6 | 20.1 |

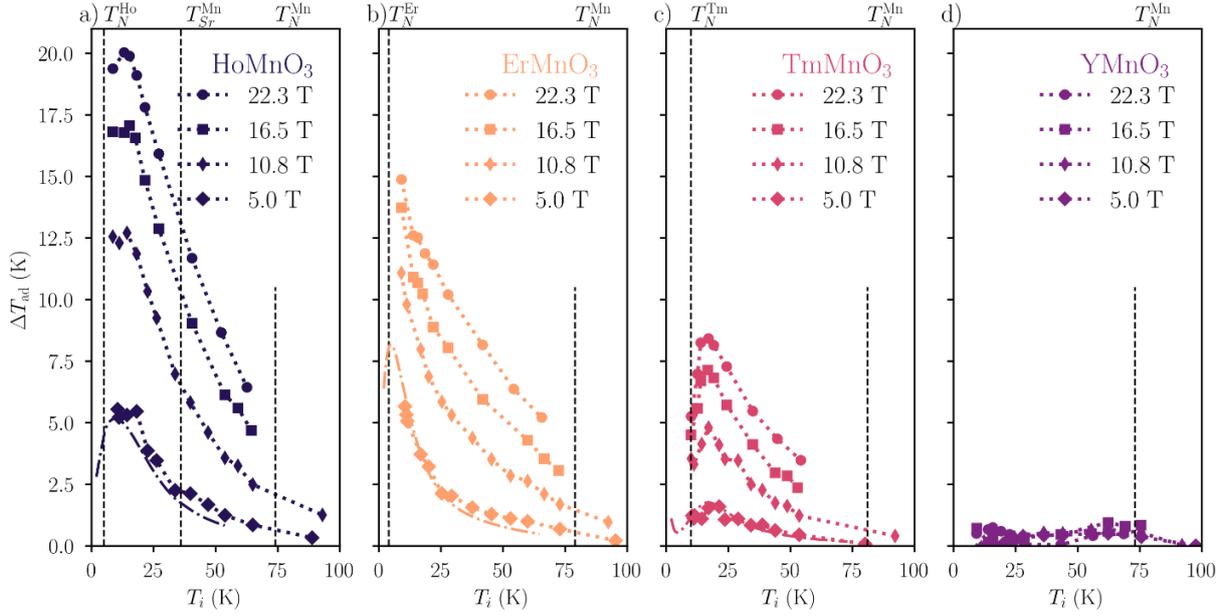

**Fig. 4:** Adiabatic temperature changes under magnetic fields. Adiabatic temperature changes are plotted as a function of the initial pulse temperature, $T_i$. The data is displayed for a) HoMnO$_3$, b) ErMnO$_3$, c) TmMnO$_3$, and d) YMnO$_3$. The data points display the adiabatic temperature changes measured directly under applied magnetic field pulses of different field strengths ranging from 5.0 T to 22.3 T; dotted lines are guides to the eye. The dashed lines are calculated adiabatic temperature changes from the entropy data measured under 5 T (Fig. 2), as described in ref. [58]. Dashed vertical lines correspond to the respective ordering temperatures determined from specific heat measurements (Fig. 2a-d).

We next move towards specific heat data recorded under applied magnetic fields up to 9 T. In agreement with previous measurements, [55] we find that the specific heat of YMnO$_3$ is independent of the applied magnetic field (Fig. 2d). In contrast, HoMnO$_3$, ErMnO$_3$ and TmMnO$_3$ show a pronounced dependency on the applied magnetic field (Fig. 2a-c). For example, with increasing magnetic field, the $T_{SR}^{\text{Ho}^{3+}}$ transition first becomes broader and then disappears, consistent with previous magnetic-field dependent specific heat studies on single crystalline HoMnO$_3$. [27] Our data further confirms that for all investigated compounds the applied magnetic field stabilizes the order of the magnetic 4f-moments, which is reflected by a field-driven increase of the transition temperature, $T_N^{R^{3+}}$. [28] From the specific heat, $C_p$, we calculate the temperature-dependent entropy change, $\Delta S_T(T,H)$, [57] for different magnetic fields, $H$, by integrating downwards from 120 K to 2 K:

$$\Delta S_T(T,H) = \int_0^T \frac{C_P(T,H) - C_P(T,H=0)}{T} dT \quad \text{Eq. 1}$$

For each of the applied magnetic fields, the calculated entropy change is plotted as a function of the temperature in Fig. 2e-h. The largest magnetic-field-induced entropy changes are observed around $T_N^{R^{3+}}$, indicating a close relation between the measured changes in entropy and the magnetic order of the rare-earth ions. This is further supported by the calculated entropy change for YMnO$_3$, which is independent of the applied magnetic field and negligible throughout the entire temperature range investigated (Fig. 2h).

To qualitatively show the magnetic-field dependence, we calculate $\Delta S_T$ and plot this difference as a function of the magnetic-field strength at $T = 13$ K in Fig. 3 (Tab. 2). Here, a clear trend of the entropy changes can be seen, with $\Delta S_T$ being most sensitive to the applied magnetic field in HoMnO$_3$, followed by ErMnO$_3$ and TmMnO$_3$.

### C. Adiabatic temperature changes

To elucidate on the origin of the observed magnetocaloric effects, we directly measure adiabatic temperature changes, $\Delta T_{\text{ad}}$, in pulsed fields up to 22.3 T (Fig. 4). For comparison, we also determined $\Delta T_{\text{ad}}$ indirectly from specific heat data, following the methodology outlined in ref. [58], for a field of 5 T. The indirectly determined values agree with the direct measurements, both in terms of absolute magnitude and temperature dependence. Interestingly, the results reveal that the adiabatic temperature change in YMnO$_3$ is largely suppressed compared to HoMnO$_3$, ErMnO$_3$, and TmMnO$_3$. The absence of a pronounced adiabatic temperature change in YMnO$_3$, together with the trend observed for HoMnO$_3$, ErMnO$_3$, and TmMnO$_3$, lead us to the conclusion that the magnetic 3d-moments of the Mn$^{3+}$ sublattice do not play an important role within the investigated magnetic field range (22.3 T) and temperature range (10 K -100 K).

This behavior at cryogenic temperature is fundamentally different from the magnetocaloric response previously discussed for other manganites at room temperature. In particular, for orthorhombic manganites, such as Ca and Sr doped LaMnO$_3$, emergent magnetocaloric effects where attribute to the Mn$^{3+}$ sublattice. [32,33] In contrast, in the cryogenic regime, we find a scaling of the effect in hexagonal manganites that strongly depends on the $R^{3+}$ ion. The maximum adiabatic temperature change for HoMnO$_3$, ErMnO$_3$, and TmMnO$_3$ occurs around $T_N^{R^{3+}}$. Furthermore, our direct measurements reveal a direct correlation between the magnetism of the $R^{3+}$ ion on the A-site and the magnetocaloric response (Tab. 2). For HoMnO$_3$, which has the largest magnetic 4f-moment, the largest magnetocaloric effect is measured with maximum adiabatic temperature changes, $\Delta T_{\text{ad}}$, of 5.2 K (5.0 T), 12.6 K (10.8 T), and 20.1 K (22.3 T). For comparison, these values correspond to 46%, 64%, and 62%, respectively, of the adiabatic temperature changes achieved with gadolinium, [59] which represent a benchmark material for magnetocaloric refrigeration at room temperature. [60]

Figure 5 shows the influence of magnetic field pulses (rise time: ~17 ms) on the adiabatic temperature change. This setting corresponds to a frequency of ~30 Hz, which is about one order of magnitude higher than the frequencies used in prototypical devices [61]. We find that the change of the adiabatic temperature with magnetic field strongly depends on the $R^{3+}$ ion, being highest for Ho$^{3+}$, followed by Er$^{3+}$, Tm$^{3+}$, and Y$^{3+}$, consistent with the magnetic-field induced entropy changes displayed in Fig. 3. The data shows an almost instantaneous response of the material's adiabatic temperature to the applied magnetic field, despite the relatively low thermal conductivity of hexagonal manganites in the temperature regime of interest (< 1 W/K/m [62]). A possible explanation is the magnetic-field dependence of the thermal conductivity, which was found to vary by up to two orders of magnitude under magnetic fields up to 14 T. [62] Importantly, the data in Fig. 5 shows the absence of pronounced hysteretic behavior, demonstrating that the magnetocaloric effect at cryogenic temperatures in hexagonal manganites is highly reversible, enabling an efficient and consistent performance over repeated thermal cycles in device applications. [63]

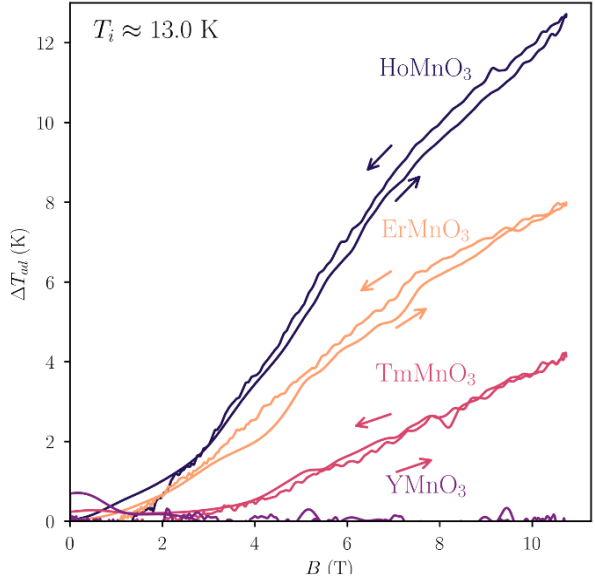

**Fig. 5:** Magnetic field-dependence of the adiabatic temperature change measured at $T_i \approx 13$ K of different rare-earth compositions at magnetic field up to 10.8 T.

### IV. Conclusions

We investigate the magnetocaloric effect and its origin in rare-earth hexagonal manganites in the cryogenic temperature regime by performing direct adiabatic temperature change measurements on polycrystalline $HoMnO_3$, $ErMnO_3$, $TmMnO_3$, and $YMnO_3$ under magnetic fields up to 22.3 T. Our results show a direct relation between the entropy change and the adiabatic temperature and the magnetic 4f-moment of the $R^{3+}$ ions, whereas contributions from the $Mn^{3+}$ sublattice were found to be irrelevant in the investigated magnetic field and temperature regime. Furthermore, the direct measurement of the adiabatic temperature changes presented in this work give robust quantitative results and prove the reversibility of the magnetocaloric effect, which previously was concluded indirectly from magnetic-field-dependent measurements. The largest magnetocaloric effect was observed for $HoMnO_3$ with $\Delta T_{ad}$ up to 20.1 K, reaching 62% of the adiabatic temperature change achieved with the benchmark material gadolinium at room-temperature.

The findings show the possibility to systematically tune the magnetocaloric response in the cryogenic regime by modifying the magnetic 4f-moments and the magnetic 4f-4f interactions in the $R^{3+}$ sublattice. This approach is of interest also for other rare-earth oxides, including rare-earth vanadates, [64] indates, [65] chromates, [66] ferrites, [67] and scandates, [68] which are increasingly recognized for their promising magnetocaloric properties at cryogenic temperatures. Finally, the multiferroic nature of hexagonal manganites within the cryogenic regime [69,70] provides a rich and so far unexplored playground for the emerging field of multicaloric cooling, [71,72] where order parameters controlled by different external fields can enhance various aspects of caloric refrigeration.

### Acknowledgements
The authors are grateful to M. Weber and M. Fiebig for helpful discussions. J.S. acknowledges financial support from the German Academic exchange service (DAAD) for a Post-Doctoral Fellowship (short-term program), NTNU Nano through the NTNU Nano Impact fund, and Tokyo Institute of Technology for funding via the World Research Hub (WRH) program. D.M. thanks NTNU for support through the Onsager Fellowship Program and the outstanding Academic Fellow Program. J.S., D.M, R.S.D., and I.H. acknowledge funding from the European Research Council under the European Union's Horizon 2020 Research and Innovation Program (grant agreement no. 863691). The authors are grateful for measurement time and direct support from the Dresden High Magnetic Field Laboratory (HLD) at Helmholtz-Zentrum Dresden-Rossendorf (HZDR), member of the European Magnet Field Laboratory (EMFL), from the EU Horizon 2020 project ISABEL (grant agreement No. 871106), and from the Clean Hydrogen Partnership and its members within the framework of the project HyLICAL (Grant No. 101101461).